\documentclass[pra,10pt,twocolumn,superscriptaddress,floatfix,showpacs]{revtex4-1}
\usepackage{subfig}
\usepackage{graphicx} 
\usepackage{epstopdf}
\usepackage[utf8]{inputenc}
\usepackage[T1]{fontenc}     
\usepackage[british]{babel}  
\usepackage[sc,osf]{mathpazo}
\usepackage[scaled=0.86]{berasans}  
\usepackage[colorlinks=true, citecolor=blue, urlcolor=blue]{hyperref}  
\usepackage[babel]{microtype}  
\usepackage{amsmath,amssymb,amsthm,bm,amsfonts,mathrsfs,bbm} 

\usepackage{xspace}  
\usepackage{pgfplots}
\usepackage{xcolor,colortbl}

\def\ba{\begin{equation}}
\def\ea{\end{equation}}
\def\bea{\begin{eqnarray}}
\def\eea{\end{eqnarray}}
\def\ben{\begin{equation*}}
\def\een{\end{equation*}}
\def\bean{\begin{eqnarray*}}
\def\eean{\end{eqnarray*}}
\def\bma{\begin{mathletters}}
\def\ema{\end{mathletters}}
\def\bi{\begin{itemize}}
\def\ei{\end{itemize}}
\newcommand{\be}{\begin{equation}}
\newcommand{\ee}{\end{equation}}

\newcommand{\ket}[1]{\ensuremath{|#1\rangle}}

\newcommand{\kommentar}[1]{}

\newcommand{\forget}[1]{}

\begin{document}

\title{Design and Simulation of an Autonomous Quantum Flying Robot Vehicle: An IBM Quantum Experience}
\author{Sudev Pradhan $^@$}
\email{sudev18@iiserbpr.ac.in}
\thanks{$@$ Those are the first authors and have equally contributed to this work}
\affiliation{Physics Department,\\ Indian Institute of Science Education and Research, Berhampur, 760010, Odisha, India}
\author{Anshuman Padhi $^@$}
\email{anshuman.padhi@niser.ac.in}
\thanks{$@$ Those are the first authors and have equally contributed to this work}
\affiliation{School of Physical Sciences,\\ National Institute of Science Education and Research, HBNI, Jatni 752050, Odisha, India}

\author{Bikash K. Behera}
\email{bikash@bikashsquantum.com}
\affiliation{Bikash's Quantum (OPC) Pvt. Ltd., Balindi, Mohanpur 741246, West Bengal, India}

\collaboration{Corresponding author: Sudev Pradhan, sudev18@iiserbpr.ac.in}
\begin{abstract}
The application of quantum computation and information in robotics has caught the attention of researchers off late. The field of robotics has always put its effort on the minimization of the space occupied by the robot, and on making the robot `smarter.  `The smartness of a robot is its sensitivity to its surroundings and the user input and its ability to react upon them desirably. Quantum phenomena in robotics make sure that the robots occupy less space and the ability of quantum computation to process the huge amount of information effectively, consequently making the robot smarter. Braitenberg vehicle is a simple circuited robot that moves according to the input that its sensors receive. Building upon that, we propose a quantum robot vehicle that is `smart' enough to understand the complex situations more than that of a simple Braitenberg vehicle and navigate itself as per the obstacles present. It can detect an obstacle-free path and can navigate itself accordingly. It also takes input from the user when there is more than one free path available. When left with no option on the ground, it can airlift itself off the ground. As these vehicles sort of `react to the surrounding conditions, this idea can be used to build artificial life and genetic algorithms, space exploration and deep-earth exploration probes, and a handy tool in defense and intelligence services. 

\end{abstract}


\maketitle

\section{Introduction \label{qrv_Sec1}}

Robotics has been a major achievement in the field of science and technology for the last half of 20th century and the 21st century \cite{qrv_Craig}. As technology progresses, the dream of robots serving in industries, production, defense, and different helpful fields have started becoming true \cite{qrv_Cai2000}. For a robot to be called more efficient, it should stand on two criteria: one, it must be intelligent, and second, it must consume less space. The rise of sensor technology and advanced computation make the robot more intelligent. And when we aim to reduce its size to the minimum possible, there comes the role of quantum effects. Studies in quantum computation have shown that it can solve various difficult classical problems more efficiently than classical computers. Quantum computation \cite{qrv_Chuang2000} also has the potential to facilitate communication with higher accuracy and is more efficient in maintaining the secrecy of quantum information. Hence, quantum computation should be used in robots to make them more efficient and occupy less space \cite{qrv_dongArXiv}. The study of robots and the engineering behind them can provide motivation for the development of powerful quantum computers.

Robot vehicles are machines that can move almost on their own. The basic purpose of robot vehicles is to make the vehicles "unmanned" i.e., no humans are there inside the vehicle to control its motion. Vehicles are equipped with sensors and computational control units. Although no user is present inside the vehicle, vehicles can still be designed to consider inputs from a distance by the user / supervisor \cite{qrv_Sand2006}. Braitenberg vehicle is a rudimentary version of a robot vehicle that can sense the light around it and move accordingly \cite{qrv_RG2019}.

Quantum robotics can trace itself back to 1998, when Benioff \cite{qrv_Benioff1998, qrv_Benioff2002} proposed the working of a quantum robot in a strictly quantum world. Then Raghuvanshi \emph{et al} showed some evidence of quantum controlled robots that used the concept of Braitenberg vehicles \cite{qrv_Raguvanshi2007, qrv_raghuvanshi2006}. The vehicle sensed its surroundings using classical light sensors(input) and moved into the environment using classical motors(output). But the sensed binary information was processed by a quantum circuit. The quantum circuit somewhat acted like the Controlling unit (CU) of the robot. The robot had three degrees of freedom, i.e. straight, left, and right.
 
Later, Mahanti \emph{et al.}  proposed the quantum circuit for a Braitenberg vehicle with an additional degree of freedom \cite{qrv_Mahanti2019}. In their setup, they made the vehicle fly off the ground in a perpendicular direction, which means it had an extra degree of freedom that was upward. Later, they went on to simulate the behavior of their proposed quantum robot deterministically. Mishra \emph{et al.} used finite automoata to propose a Quantum Braitenberg vehicle. and were successful in including external control to its movement \cite{qrv_mishra2019}. A month later, the same research group proposed an algorithm in which complex desired behaviors such as the addition of an additional degree of freedom could be incorporated into the vehicle system \cite{qrv_mishra1.2019}. 

In the current work, we build on the quantum Braitenberg vehicles and propose a smarter vehicle, which is more sensitive to its surroundings than the previously discussed models. In this proposal, instead of using light sensors as in the case of Braitenberg vehicles, here we use IR sensors for better detection of obstacles; since light as a sensation can be misleading and the vehicle's motion might get manipulated by. It has all 4 degrees of freedom when on the ground (i.e., forward, right, left, and backward). Another interesting feature of the vehicle is, every time it senses more than one possible direction which the robot can traverse (meaning that more than one direction has no obstacles), it asks the user to specify the direction which it should take. Later, as the user specifies, the robots proceed according to the instruction. The robot has another feature, that is, its ability to elevate itself up in the air when there is no path available to traverse on the ground. For the above functions and features of the robot, we develop a quantum circuit and realize on the IBM Q platform, and validate all the results.

\begin{figure}[!tbp]
  \centering
  \subfloat[A classical Braitenberg Vehicle]{\includegraphics[width=0.4\textwidth]{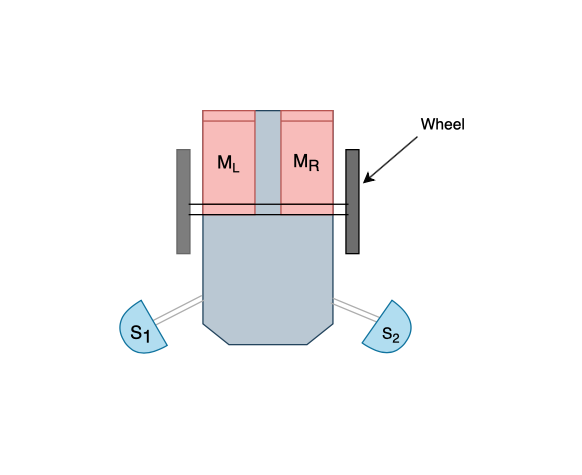}\label{fig:f1}}
  \hfill
  \subfloat[A Quantum robot vehicle]{\includegraphics[width=0.4\textwidth]{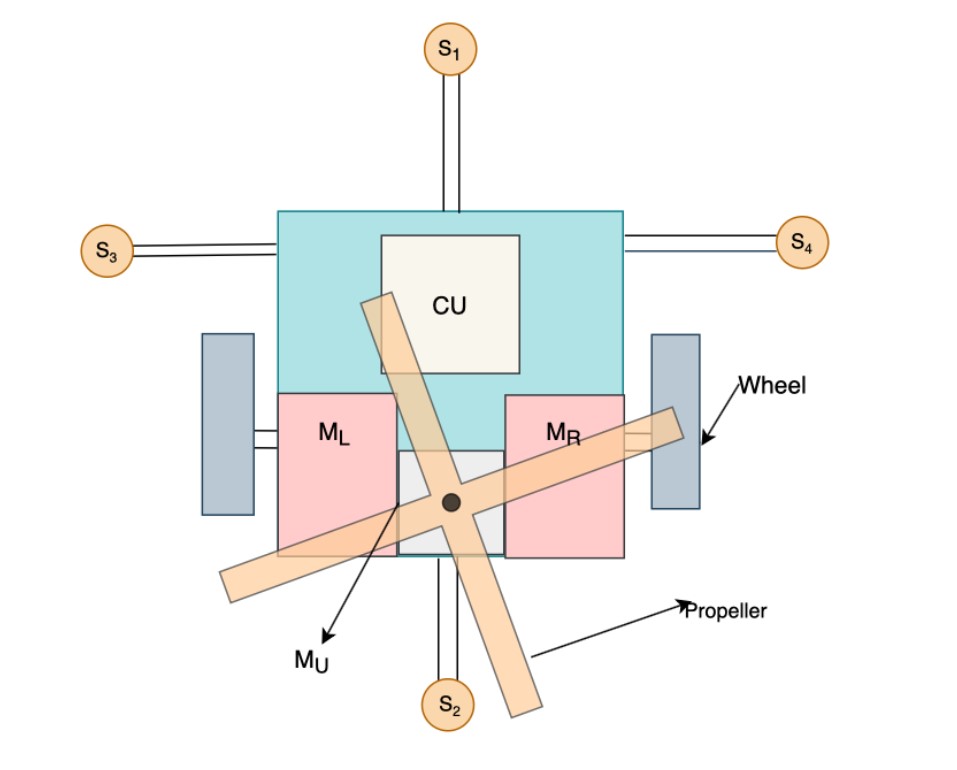}\label{fig:f2}}
  \caption{textbf{The schematic top view of the classical Braitenberg vehicle and the proposed quantum robotic vehicle.} Figure (a) The light sensors are named as $S_{1}$ and $S_{2}$ are connected to the motors $M_{L}$ and $M_{R}$ respectively. Both the motors are connected to the respective wheels. Figure (b) corresponds to the schematic top-view of the proposed quantum robot vehicle. Instead of having 2 light sensors like the classical one, it is equipped with 4 IR sensors $S_{1}$, $S_{2}$, $S_{3}$, $S_{4}$ at the front, back, left and right respectively. Inputs from these sensors is fed into the Controlling Unit (CU) of the robot and the output of the CU lets the motors move accordingly. Motors $M_{L}$ and $M_{R}$ are connected to the wheels in the respective sides. The motor $M_{U}$ is connected to the propeller attached to the vehicle. When $M_{U}$ is active, the propeller works and the vehicle takes up from the ground.}
  \label{qrv_Fig1}
\end{figure}
Here, we design the required quantum circuit and simulate it by using IBM Quantum Experience. IBM has recently developed various prototypes of quantum processors and has made it available through a free web-based, namely, IBM Quantum Experience. Researchers have used it to their strengths and have been able to demonstrate and run a variety of experiments involving quantum computation, e.g. \cite{qrv_IBMQE1, qrv_IBMQE2, qrv_IBMQE3, qrv_IBMQE4, qrv_IBMQE5, qrv_IBMQE6, qrv_IBMQE7,qrv_IBMQE8, qrv_IBMQE9, qrv_IBMQE10, qrv_IBMQE11, qrv_IBMQE12,qrv_ManabputraQIP2019}. In this experiment, we use IBM Quantum Experience's; `qasm simulator' and `Jupyter notebook' to construct and simulate the circuit.

We organize this paper as follows. Section \ref{qrv_Sec2} describes the theoretical scheme of the robot vehicle that we tend to propose. Section \ref{qrv_Sec3} proposes the logic behind the design of the quantum circuit and the various features it will provide to the vehicle. Section \ref{qrv_Sec4} describes how the simulation of the proposed logic was performed and the result of it. Section \ref{qrv_Sec5} explains the applications of this vehicle in real life. Finally, we conclude this article by citing the future implications of the proposal in Section \ref{qrv_Sec6}.

\section{SCHEME OF THE ROBOT VEHICLE \label{qrv_Sec2}}

Braitenberg vehicles are made up of light sensors and wheels \cite{qrv_MIT1992}. The fundamental plan behind the vehicle is that, supported the device input, the wheels mechanically move in a manner accordingly, therefore giving rise to differing kinds of motion to the wheels which ultimately move the vehicle. Depending upon the type of connection between the sensors and the wheels, the movements of the vehicle can be specified when there is a certain type of external surrounding. The vehicle can also change its motion according to the changing external environment.

A schematic diagram of a simple Braitenberg vehicle is delineated in Fig. \ref{qrv_Fig1}(a). As observed, the two light sensors $S_{1}$ and $S_{2}$ sense the intensity of the incident light. They are connected to the two motors $M_{L}$ and $M_{R}$ in such a way that when a light sensor senses some light, then the motor connected to it starts rotating and the motor is turned off in the absence of light. When the robot is exposed to high light intensity, both the sensors sense light; hence, both the motors get turned on and as a result, the robot moves forward at full speed. In the case where only the left light sensor $S_{1}$ senses some light, only the left motor $M_{L}$ starts rotating and $M_{R}$ does not and that gives the motor a rightward motor moves towards the right-hand side. It can be attributed to the formation of an instantaneous axis of rotation around the right wheel, since it is at rest and the body turns around it, thus having a motion towards right. The same goes for the case of light coming from the right-hand side, in which case the motor turns to the left. This behavior of the robot to deviate away from intense light sources and obstacles can be interpreted as `fear' or `shyness'. Similarly, we can make the robot drive itself towards light by changing the connections in the circuit, and thus the nature of that sort of Braitenberg vehicle will be opposite to that of ``fear" or ``shyness" \cite{qrv_MIT1992}.

The original classical Braitenberg vehicle's motion is determined by the presence or absence of light on the left and right sides, so it has a particular disadvantage that its forward motion is dependent on the intensity inputs of the left and right sensors. In these cases, the motion is very much dependent on the intensity of the lights, which at times can be unpredictable. Also, using intensity as input, it may collide with some obstruction that is bright enough and cause itself some damage. It also has no means of moving back.

In our work, we propose a modified Braitenberg vehicle in Fig. \ref{qrv_Fig1}(b) that senses the obstacles in its path rather than detecting the light intensity. For this purpose, it is equipped with 4 IR sensors, each at the front, back, left and right. The sensor inputs give us an idea on which side there are obstructions to the robot, and on which side the path is clear. The inputs are then fed into the quantum circuit we have designed. Out of 4 directions, if exactly one direction is obstacle-free, the circuit will give output to the motors to make the movement possible in that direction. If more than 1 direction is available to be traveled upon, then the circuit lets the robot ask the user to specify which direction is to be traveled. Along with the $M_{L}$ and $M_{R}$ motors on each side, an additional motor $M_{U}$ is attached to the propeller of the robot. When there are obstructions from all 4 sides, the propeller will rotate to let the robot take off from the ground and fly upward.

When there are obstructions in front, left and right, the robot will also be able to move backward.

\section{Quantum Scheme \label{qrv_Sec3}}

The proposed robot has 4 sensors $S_{1}$, $S_{2}$, $S_{3}$, $S_{4}$ each at the front, back, left, and right, respectively. Each sensor is assigned 1 qubit each. We denote the state $\ket{0}$ when the sensor detects an obstacle and $\ket{1}$ when the sensor detects no obstacle on its front. The input of all 4 sensors will say which are the directions around the vehicle that are devoid of any obstacle (Table \ref{Input}). There are 3 motors $M_{L}$, $M_{R}$ and $M_{U}$ each attached to the left wheel, right wheel, and the propeller at the top, respectively. $M_{L}$ and $M_{R}$ are assigned 2 qubits each. $\ket{00}$ state of these motors indicates that the wheel is not rotating at all, $\ket{10}$ means that the wheel is moving forward, and $\ket{01}$ indicates that the wheel is moving backward (Table \ref{Output}).

$M_{U}$ is attached to the propeller and is assigned only 1 qubit. The $\ket{0}$ state indicates that the propeller is not rotating, and the $\ket{1}$ state indicates that the propeller is rotating. When the propeller is rotating (i.e. $M_{U}$ has a $\ket{1}$ state) the robot takes off from the ground.

\begin{table}[]
\centering
\begin{tabular}{|c|c|}
\hline
 Input received by  & Remarks  \\
  all sensors $\ket{ S_{1} S_{2} S_{3} S_{4}}$ &  \\
\hline
\hline
$\ket{0000}$ & Obstruction at all 4 sides \\ \hline
$\ket{0001}$ &  Right path clear\\ \hline
$\ket{0010}$ &  Left path clear\\ \hline
$\ket{0100}$ &  Back path clear\\ \hline
$\ket{1000}$ &  Front path clear\\ \hline
$\ket{1100}$ & \\
$\ket{1010}$ & \\
$\ket{1001}$ & \\
$\ket{0110}$ & \\
$\ket{0101}$ & Cleared path in more \\
$\ket{0011}$ & than one direction \\
$\ket{1110}$ & \\
$\ket{1101}$ & \\
$\ket{1011}$ & \\
$\ket{0111}$ & \\
$\ket{1111}$ & \\
\hline         
\end{tabular}

\caption{\textbf{All possible inputs and their interpretations}}
\label{Input}
\end{table}

\begin{table}[]
\centering
\begin{tabular}{|c|c|}
\hline
 State of the qubits  & Corresponding action  \\
 attached to the motor & on the wheel attached to it \\
\hline
$\ket{00}$ & No rotation \\
$\ket{01}$ & Wheel rotates and moves backwards \\
$\ket{10}$ & Wheel rotates and moves forward \\
\hline         
\end{tabular}
\caption{\textbf{State of the qubits assigned to the motor and the corresponding output on the wheel attached to that motor}}
\label{Output}
\end{table}

The quantum circuit is designed such that, when the inputs collectively exhibit a $\ket{1000}$ state, which implies there are obstructions in left, right and back and a free path in the front; both the $M_{L}$ and $M_{R}$ have a forward motion i.e. both the motors have $\ket{10}$ states. Similarly, when the sensor input gives a $\ket{0001}$ state, it indicates a vacancy in the backside of the robot, and thus the output of the circuit will result in both $M_{L}$ and $M_{R}$ having a backward motion, i.e. both have a $\ket{01}$ state. When the input reads $\ket{0010}$, that is, there is a vacancy only on the left and obstructions on the rest of all directions; to make the robot turn left, the left wheel is made stationary and the right wheel is made to move forward, and as discussed earlier, it makes the vehicle turn left. Hence, the design of the quantum circuit is such that, when receiving an input of $\ket{0010}$, $M_{L}$ shows an output $\ket{00}$, while $M_{R}$ shows an output $\ket{10}$. Similarly, upon receiving an input $\ket{0001}$ (vacancy on the right), the right motor $M_{R}$ must be stationary, i.e., state $\ket{00}$, and the left motor $M_{L}$ must have a value $\ket{10}$ state.

When the inputs read $\ket{0000}$ i.e. there are obstacles on all 4 sides and there is no path for the robot to move on the ground; the circuit renders the $M_{U}$ qubit to be in the $\ket{1}$ state, the propellor starts rotating and the robot takes off from the ground.

There may be cases in which there can be more than 1 direction available to be traveled upon. This implies that the circuit input shows more than one $\ket{1}$ state out of the 4 sensors (the inputs like $\ket{0011}$, $\ket{1001}$, $\ket{1110}$, etc.).  In those cases, we aim for the robot to stand still, without moving anywhere on its own, and asks the user to specify which direction to move. Hence we define one `ASK' qubit in the circuit. When in the $\ket{0}$ state, it does not ask and keeps on moving as per the motor qubits. But when in the $\ket{1}$ state, it asks the user to specify the direction. The user then specifies the direction through the quantum computer control it has, and the robot moves in that direction. Taking into account all these, we should devise a quantum circuit such that it satisfies Table \ref{Truth Table}, since it describes how the circuit should behave in terms of its input and output, so that the desired property is obtained.

\begin{table}[]
\centering
\begin{tabular}{|c|c|c|c|c|l|}
\hline
 Input received   & Left & Right & Propellor & ASK & Action of    \\
  by all sensors  & Motor & Right & Motor & qubit & the vehicle\\
  $\ket{S_{1} S_{2} S_{3} S_{4}}$ & ML($\ket{ab}$) & MR($\ket{ab}$) & & & \\
\hline
\hline
$\ket{0000}$ & $\ket{00}$ & $\ket{00}$ & $\ket{1}$ & $\ket{0}$ & flies up \\ \hline
$\ket{0001}$ & $\ket{10}$ & $\ket{00}$ & $\ket{0}$ & $\ket{0}$ &  right\\ \hline
$\ket{0010}$ & $\ket{00}$ & $\ket{10}$ & $\ket{0}$ & $\ket{0}$ & left\\ \hline
$\ket{0100}$ & $\ket{01}$ & $\ket{01}$ & $\ket{0}$ & $\ket{0}$ &  back\\ \hline
$\ket{1000}$ & $\ket{10}$ & $\ket{10}$ & $\ket{0}$ & $\ket{0}$ & front\\\hline
$\ket{1100}$ & $\ket{00}$ & $\ket{00}$ & $\ket{0}$ & $\ket{1}$ & \\
$\ket{1010}$ & $\ket{00}$ & $\ket{00}$ & $\ket{0}$ & $\ket{1}$ &\\
$\ket{1001}$ & $\ket{00}$ & $\ket{00}$ & $\ket{0}$ & $\ket{1}$ &\\
$\ket{0110}$ & $\ket{00}$ & $\ket{00}$ & $\ket{0}$ & $\ket{1}$ &\\
$\ket{0101}$ & $\ket{00}$ & $\ket{00}$ & $\ket{0}$ & $\ket{1}$ & no move\\
$\ket{0011}$ & $\ket{00}$ & $\ket{00}$ & $\ket{0}$ & $\ket{1}$ & asks  \\
$\ket{1110}$ & $\ket{00}$ & $\ket{00}$ & $\ket{0}$ & $\ket{1}$ & \\
$\ket{1101}$ & $\ket{00}$ & $\ket{00}$ & $\ket{0}$ & $\ket{1}$ &\\
$\ket{1011}$ & $ \ket{00}$ & $ \ket{00}$ & $\ket{0}$ & $\ket{1}$ &\\
$\ket{0111}$ & $\ket{00}$ & $\ket{00}$ & $\ket{0}$ & $\ket{1}$ &\\
$\ket{1111}$ & $\ket{00}$ & $\ket{00}$ & $\ket{0}$ & $\ket{1}$ &  \\
\hline         
\end{tabular}
\caption{\textbf{Details of the robot behaviour upon the sensation of obstruction.}}
\label{Truth Table}
\end{table} 

From the table, we can observe that we have 16 types of cases for inputs (as we have total 4 sensors). We prepare a conditional circuit with 16 segments. Each segment is prepared from a four-control Toffoli gate, conditioned to receive only a specific input of 16. When that input is available, only that segment out of all 16 works and only the NOT gate (s) attached to that Toffoli gate operate accordingly on the qubits to be measured, and thus we obtain our desirable outputs.

\begin{figure}[]
    \centering
   \includegraphics[scale=0.2]{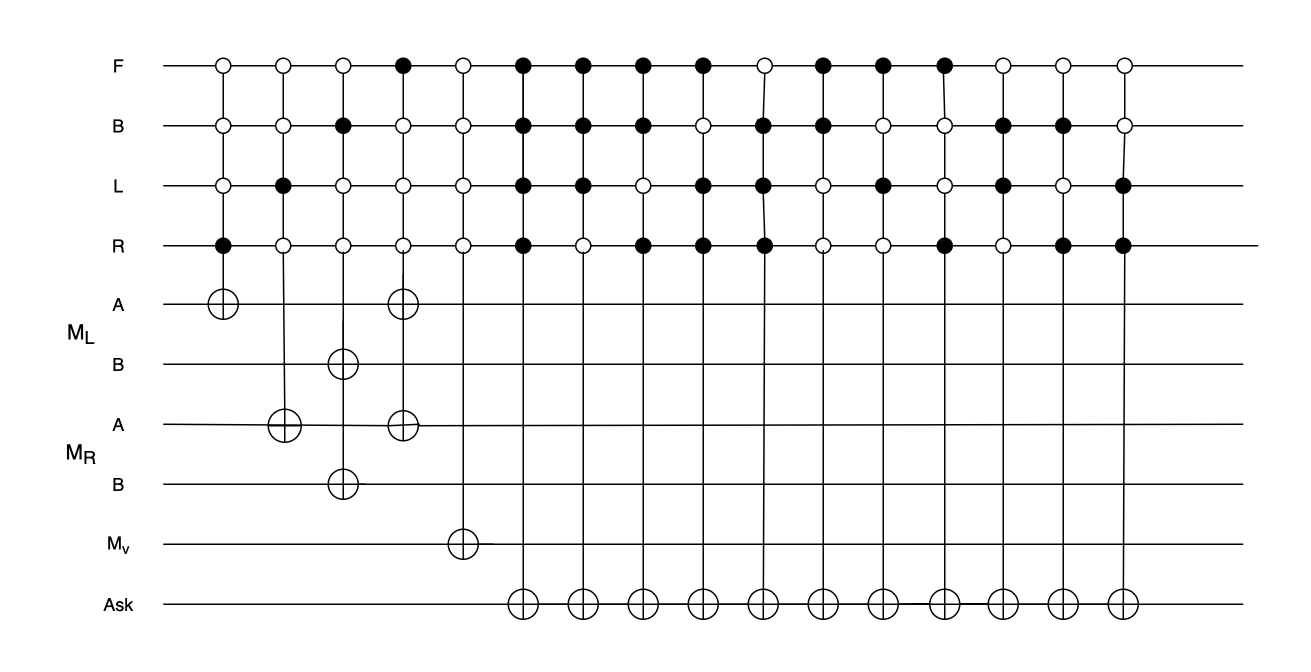}
    \caption{\textbf{Quantum circuit for implementing the proposed protocol of the robot vehicle.} The Qubits F, B, L, and R imply the input from the sensors $S_{1}$, $S_{2}$, $S_{3}$, $S_{4}$ respectively. The motors $M_{L}$ and  $M_{R}$ are assigned two qubits each, namely `A' and `B'. The motor $M_{U}$ is assigned 1 qubit and the robot is assigned a qubit to ASK the user to clarify the direction. The qubits of the motors $M_{L}$, $M_{R}$, $M_{U}$ and the ASK qubit are measured as output. It should be noted that there are 16 vertical segments (Four-controlled Toffoli gates) in it, each representing one of the total 16 types of possible inputs.}
    \label{Logic Circuit}
\end{figure}

From this table, we can propose a quantum circuit in Fig. \ref{Logic Circuit}. Suppose, for the input $\ket{0001}$, we can see that only the first segment (i.e., the first Toffoli gate) gets activated, and now as per Fig. \ref{Logic Circuit}, the corresponding NOT gate operates on the first qubit of $M_{L}$ (i.e., $M_{L}^{A}$) and the state of the left motor $M_{L}$ changes from $\ket{00}$ to $\ket{11}$. Similarly, for each segment (i.e. each possibility of input) the NOT gates are implemented as targets on different qubits to obtain the desired output.

\section{QUANTUM CIRCUIT AND SIMULATION \label{qrv_Sec4}}
\subsection{Working and Decomposition of four-control-one target and four-control-two-target Toffoli gate-}
\begin{figure}
    \centering
   \includegraphics[height=5cm , width=9cm]{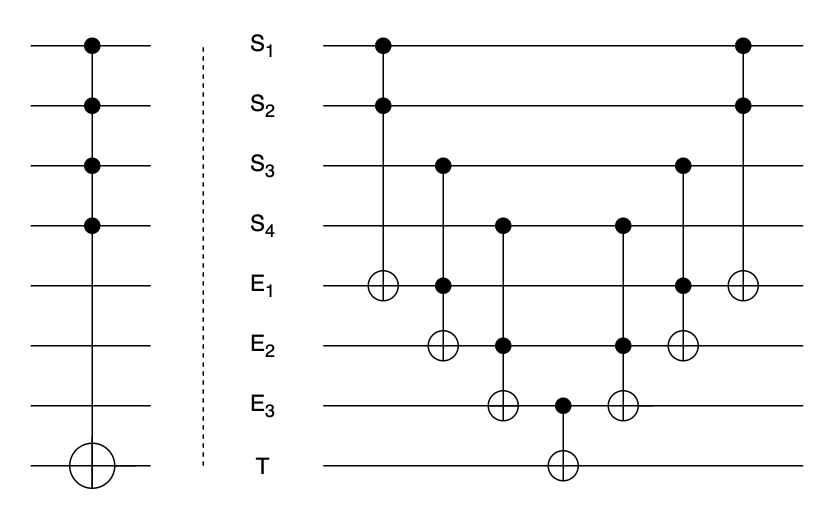}
    \caption{\textbf{Equivalent quantum circuit for four-control-one-target ($C^{4}(X_{5})$) Toffoli gate.} The four-control-one-target Toffoli gate is fed with 3 extra qubits, which act as a buffer when the Four-control-one-target Toffoli gate is decomposed into two-control-one-target Toffoli gates. It is to be observed that the extra qubits undergo two NOT gates in total, hence they will always show $\ket{0}$ as output.}
    \label{Decompose}
\end{figure}

The operation of a 4-control Toffoli gate with one target ($C^{4}(X_{5})$) is as follows. There are 4 control qubits and one NOT gate. When the control qubits are satisfied, the NOT gate flips the target qubit. However, since it is difficult to prepare a four-control-one-target Toffoli gate, we break it into a combination of two-control-one-target and one-control-one-target Toffoli gates (refer to Fig. \ref{Decompose}). But to do so, we additionally will require three more qubits, named Extra Qubits (Ex 1, Ex 2, Ex 3). The sole purpose of the extra qubits is to facilitate the breakdown of a four-control Toffoli gate into a combination of two-control Toffoli gates. 7 two-control Toffoli gates and 1 one-control Toffoli gate are used to denote a four-control, one-target Toffoli gate.

\subsection{Circuit Diagram and Simulation result-}
\begin{figure}
    \centering
   \includegraphics[width=8cm,height=22cm]{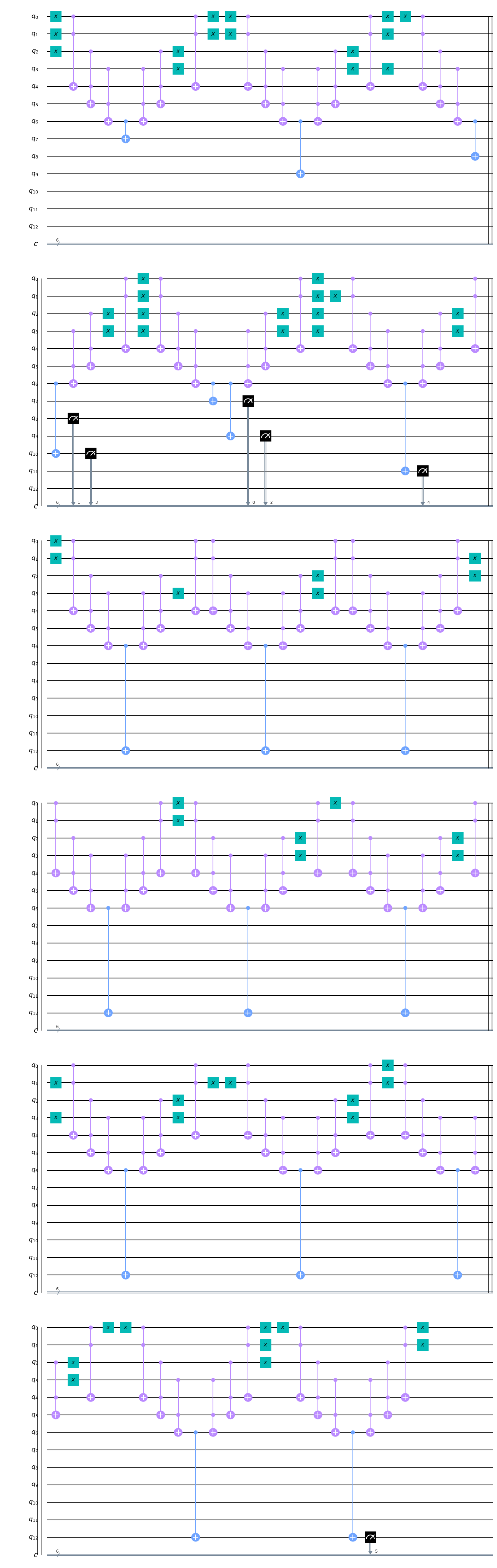}
    \caption{\textbf{Construction of the Quantum Circuit using IBM Q Experience}}
    \label{IBMQ}
\end{figure}

We use Fig. \ref{IBMQ} for the proposed experiment. The first 4 qubits (q[0] to q[3]) are used as the auxiliary qubits. They represent the inputs received by the sensors at Front, Back, Left, and Right, respectively. The next 3 qubits (q[4]-q[6]) are the `extra' qubits discussed in the previous subsection. They will be used as buffer to accommodate the decomposed four-control-one-target Toffoli gate. Qubits q[7] and q[8] correspond to the left motor, which has been assigned two qubits. Similarly, q[9] and q[10] correspond to the two qubits of the motor at right. In this proposed circuit, q[11] is the qubit attached with the motor connected to the propeller. The final qubit q[12] is associated with the robot asking for the specification of direction to its user. While the qubits q[0]-q[3] are the inputs to the circuit, the qubits q[7]-q[12] are taken for measurement. After this we use the appropriate gates, as discussed in the previous subsection, to prepare the IBM quantum circuit in Fig.\ref{IBMQ}, ready for simulation.
\begin{table}[]
\centering
\begin{tabular}{|c|c|c|}
\hline
 Input  & Output & Action  \\
  $\ket{S_{1}S_{2}S_{3}S_{4}}$ & $\ket{(Ask)M_{U}M_{R}^{B}M_{R}^{A}M_{L}^{B}M_{L}^{A}}$ & \\
\hline
\hline
$\ket{0000}$ & $\ket{010000}$ & Flies Up \\ \hline
$\ket{0001}$ & $\ket{000001}$ & Takes right\\ \hline
$\ket{0010}$ & $\ket{000100}$ & Takes left \\ \hline
$\ket{0100}$ & $\ket{001010}$ &  Moves back\\ \hline
$\ket{1000}$ & $\ket{000101}$ &  Moves forward\\ \hline
$\ket{1100}$ & $\ket{100000}$ &  \\
$\ket{1010}$ & $\ket{100000}$ &  \\
$\ket{1001}$ & $\ket{100000}$ &  \\
$\ket{0101}$ & $\ket{100000}$ &  Doesn't move\\
$\ket{0110}$ & $\ket{100000}$ & and asks \\
$\ket{0011}$ & $\ket{100000}$ &  the user\\
$\ket{1110}$ & $\ket{100000}$ &  to specify\\
$\ket{1101}$ & $\ket{100000}$ &  direction\\
$\ket{1011}$ & $\ket{100000}$ &  \\
$\ket{0111}$ & $\ket{100000}$ &  \\
$\ket{1111}$ & $\ket{100000}$ &  \\
\hline         
\end{tabular}

\caption{\textbf{Result of Simulation of Quantum Circuit} }
\label{Simulation}
\end{table}

When the ancillary qubits were varied and all 16 possible cases were tested, the obtained result was tabulated in Table \ref{Simulation}.

\section{ADVANTAGE and APPLICATIONS \label{qrv_Sec5}}
The proposed and simulated quantum robot vehicle has more degrees of freedom than the previously proposed ones. It can autonomously move to its left, right, back and forth, subject to the condition that there is no obstacle in that path. It is smart enough to handle complex situations like the presence of more than one possible path to traverse. In conditions like these, it will always ask the user to further specify the direction. When left without options on the ground, it can also take off from the ground using a propellor. The drone mechanism can be later added to make it move in all directions in the air as well. 

The application of quantum phenomena ensures that it has a minimized size and reduced weight. As it can also be controlled by the user when necessary, it can go to places that are unreachable or hazardous to humans. This can open up the possibility of a great field of scientific inquiry. Like the one we did, quantum robots can handle complex situations effectively, without much effort going onto it. Robots with a full-fledged quantum control unit can also ensure that the secrecy of the information obtained remains intact. The use of quantum computation and information in the future can make it fit enough to be used as rovers for planetary and deep-earth explorations. It can also serve intelligence and defense agencies in protecting their territory.

\section{CONCLUSION \label{qrv_Sec6}}
The application of quantum computing in robotics can be a boon for technological advancements in the 21st century. It started with Benioff giving an idea about a quantum robot in a purely quantum domain. Scientists have taken inspiration from this and have been able to realize the potential that lies in the application of quantum phenomena in the field of robotics. It can be seen that quantum robots can solve complex problems more efficiently than classical computers.  We can improve the setup further by using quantum sensors, and it can be made faster by using parallel computation and learning of quantum algorithms.

Braitenberg vehicles have been used for mimicking animal movements, and thus setting up a whole field of scientific inquiry on artificial life, genetic algorithms etc. \cite{qrv_atificial2018}. Using quantum computation can effectively analyze the complex behaviors that animals show, such as ‘fear’ and ‘attraction’  \cite{qrv_RG2019}; which governs their movements and actions. This can later lead to the incorporation of various other emotions, and a quantum emotional robot can be thought of \cite{qrv_MIT1992, qrv_Mahanti2019}.

We used quantum computation to make the robot navigate. Similarly, there is a wide scope for quantum computation and information in robot action planning and vision, which can further speed up the processes involved in robots, making them compact and efficient.

\acknowledgments
A.P. and S.P. would like to thank Bikash's Quantum (OPC) Pvt. Ltd. for providing hospitality during the course of this project. B.K.B. acknowledges the prestigious Prime Minister's Research Fellowship awarded by DST, Govt. India. We acknowledge the support of IBM Quantum Experience for producing the basic circuits. The views expressed are those of the authors and do not reflect the official policy of IBM or IBM Quantum Experience team.
\section*{Data Availability}

Data will be made available on reasonable request.
\section*{Competing interests}
The authors declare that they have no competing financial and nonfinancial interests. Readers are welcome to comment on the online version of the paper. Correspondence and requests for materials should be addressed to S.P. (sudev18@iiserbpr.ac.in).

\end{document}